# A revisit to phonon-phonon scattering in single-layer graphene

Xiaokun Gu[1#], Zheyong Fan[2], Hua Bao[3], C. Y. Zhao[1*]

[1]Institute of Engineering Thermophysics, School of Mechanical Engineering, Shanghai Jiao Tong University, Shanghai 200240, China

[2] QTF Centre of Excellence, Department of Applied Physics, Aalto University, FI-00076 Aalto, Finland

[3]University of Michigan-Shanghai Jiao Tong University Joint Institute, Shanghai Jiao Tong University, Shanghai 200240, China

[#]Email: Xiaokun.Gu@sjtu.edu.cn

[*]Corresponding author. Email: Changying.Zhao@sjtu.edu.cn

**Abstract**

Understanding the mechanisms of thermal conduction in graphene is a long-lasting research topic, due to its high thermal conductivity. Peierls-Boltzmann transport equation (PBTE) based studies have revealed many unique phonon transport properties in graphene, but most previous works only considered three-phonon scatterings and relied on interatomic force constants (IFCs) extracted at 0 K. In this paper, we explore the roles of four-phonon scatterings and the temperature dependent IFCs on phonon transport in graphene through our PBTE calculations. We demonstrate that the strength of four-phonon scatterings would be severely overestimated by using the IFCs extracted at 0 K compared with those corresponding to a finite temperature, and four-phonon scatterings are found to significantly reduce the thermal conductivity of graphene even at room temperature. In order to reproduce the prediction from molecular dynamics simulations, phonon frequency broadening has to be taken into account when determining the phonon scattering rates. Our study elucidates the phonon transport properties of graphene at finite temperatures, and could be extended to other crystalline materials.

I. Introduction

Inspired by the reported ultrahigh thermal conductivity from measurements [1, 2], phonon transport properties of graphene were intensively investigated due to its potential for thermal management [3, 4]. Quite a few theories that are originated from various pictures of phonons [5], such as molecular dynamics (MD) simulations [6-11], atomistic Green's function [12, 13] and Peierls-Boltzmann transport equation (PBTE) based methods [14-20], have been applied to study the phonon transport in graphene and related nanostructures. Since the PBTE approach could be easily integrated with first-principles [21-24], making it be of the predictive power on thermal conductivity, and provide the detailed information on phonon scattering and phonon mean free path, it has become a routine to explore the thermal transport properties of crystalline materials. Fruitful results from the PBTE approach have been obtained, which deepen our understanding for thermal transport in graphene and other two-dimensional materials [25, 26].

Despite considerable progress having been made, there are several concerns that might affect the usage of the PBTE approach on predicting the thermal properties of graphene. Firstly, in most previous studies on graphene, three-phonon scatterings are treated as the main origin of inelastic phonon scatterings and higher-order phonon scatterings are ignored. However, recent studies have shown that four-phonon scatterings have to be taken into account in the PBTE calculations for some bulk materials, such as cubic silicon [27], hexagonal silicon [28], boron arsenide [29], even at room temperature, otherwise the predicted thermal conductivity would be overestimated considerably. Very recently, Feng and Ruan [30] calculated the thermal conductivity of graphene whose interatomic interaction is described by the optimized Tersoff potential [31] by considering both three-phonon and four-phonon scatterings. They reported that the thermal conductivity of graphene is around 800 W/mK, far below the previously reported simulation data (~ 3500 W/mK) from PBTE calculations where four-phonon scatterings are not included [18, 31]. This work seems to confirm the importance of four-phonon scatterings in graphene, but the obtained thermal conductivity value is substantially lower than the results from MD simulations using the same empirical potential [6-9, 11], where the four-phonon scatterings are naturally captured. The origins of the discrepancy of the two numerical methods have not been fully understood yet. One possible example explanation is that the phonons in MD simulations do not follow the Bose-Einstein distribution but are fully excited.

Secondly, interatomic force constants (IFCs) used in PBTE studies on graphene were usually obtained at the static limit (0 K), and the temperature effects on these IFCs were not taken into account [14, 16, 18, 20]. One example to stress the importance of the temperature dependent IFCs on phonon transport is crystalline materials with structural phase transition at high temperatures. Imaginary frequencies, a sign of instability, usually occur if using the interatomic force constants extracted at 0 K to compute the phonon dispersions of the high-temperature phases [32, 33]. Only by taking the temperature dependence on IFCs into account, the vibrational properties of high-temperature phases could be correctly captured [34]. Even if no structural phase transition happens, the anhamonic interactions might result in phonon shifts, which can be regarded as an outcome of temperature dependent force constants and might have some impacts on phonon scatterings as well as thermal conductivity [35, 36]. The effects of temperature dependent IFCs on thermal conduction in graphene have not been reorganized.

Thirdly, when calculating the phonon scattering rates appearing in the PBTE, one has to employ the Dirac delta function to ensure energy conservation condition before and after the phonon scattering [37]. Several numerical schemes have been proposed to treat the Dirac delta function when integrating the first Brillouin zone. The representative approaches include analytical methods, in which the first Brillouin zone is divided into tetrahedrons [38] or stripes [39] and the integration in each small region is analytically computed, and finite-breath function method, where the Dirac delta function is replaced by a Gaussian/Lorentzian function with either fixed [40] or adaptive smearing parameters [23, 41]. For the finite-breath function method with adaptive smearing parameters, Li *et al*. [23] suggested to connect the smearing parameters with the group velocities of phonons that take part in the scattering, while Turney *et al*. [41] linked those with phonon linewidths. A more detailed analysis to clarify the distinction among these approaches is highly required in order to accurately predict the phonon transport using the PBTE formalism.

In this paper, we perform PBTE calculations to determine the thermal conductivity of graphene modeled by the optimized Tersoff potential. Three-phonon scatterings, four-phonon scatterings, and phonon-boundary scatterings are fully taken into account in the phonon transport modeling. While most PBTE studies used IFCs from first-principles as inputs and tried to match the measured experimental data, we employ the empirical potential to generate the interatomic force constants. This is because the thermal conductivity measurements on two-dimensional materials are still quite challenging and there is still no consensus on the accuracy of the existing measurement methods [42]. Instead, using the empirical potential, we could compare the results from PBTE studies with the thermal conductivity predictions made by MD simulations directly, and identify the influence of four-phonon scatterings and temperature dependent IFCs on phonon transport in graphene. In addition, we also employ different schemes to approximate the Dirac delta function for energy conservation, and the validity of these methods is discussed. To fairly compare the results from the MD and PBTE calculations, both classical and quantum phonon population functions are used in our calculations.

## II. Theoretical Methods
### A. Interatomic force constants

The vibrational properties of a material are determined by the Hamiltonian, which is expressed as [43]

$$H = T + V$$

$$T = \sum_i T_i$$

$$V = E_0 + \frac{1}{2!}\sum_{ij}\sum_{\alpha\beta} \phi_{ij}^{\alpha\beta} u_i^\alpha u_j^\beta + \frac{1}{3!}\sum_{ijk}\sum_{\alpha\beta\gamma} \psi_{ijk}^{\alpha\beta\gamma} u_i^\alpha u_j^\beta u_k^\gamma$$

$$+ \frac{1}{4!}\sum_{ijkl}\sum_{\alpha\beta\gamma\theta} \chi_{ijkl}^{\alpha\beta\gamma\theta} u_i^\alpha u_j^\beta u_k^\gamma u_l^\theta + \cdots . \tag{1}$$

where $T$ and $V$ are the kinetic and potential energies; the pair $(i, \alpha)$ means the degree of freedom in the material corresponding to the $\alpha$ direction of atom $i$; $u$ is the atomic displacement away from the equilibrium position; $\phi$, $\psi$ and $\chi$ are the so-called second-order harmonic, third-order and fourth-order anharmonic force constants; $E_0$ is the potential energy when $u = 0$. The concept of phonons is based on the harmonic oscillators, which assumes the anharmonic terms (the last two terms) in Eq. (1) are much smaller than the harmonic term (the second term on the right-hand side) of Eq. (1). The second-order harmonic force constants determine many phonon properties, such as the phonon dispersion, phonon group velocity and heat capacity of a specific mode. Based on the perturbation theory [43], the anharmonic terms lead to phonon scatterings, which results in the resistance to the heat flow. The strength of phonon-phonon scatterings could be calculated using the information of anharmonic force constants. The atomic forces are also related to the IFCs through [44]

$$F_i^\alpha = -\frac{\partial V}{\partial u_i^\alpha} = -\sum_j \sum_\beta \phi_{ij}^{\alpha\beta} u_j^\beta - \frac{1}{2!}\sum_{jk}\sum_{\beta\gamma} \psi_{ijk}^{\alpha\beta\gamma} u_j^\beta u_k^\gamma - \frac{1}{3!}\sum_{jkl}\sum_{\beta\gamma\theta} \chi_{ijkl}^{\alpha\beta\gamma\theta} u_j^\beta u_k^\gamma u_l^\theta + \cdots \quad (2)$$

Mathematically, $\phi$, $\psi$ and $\chi$ are regarded as the negative first-order, second-order and third-order derivatives of atomic force with respect to atomic displacements. To obtain the IFCs, one, two or three atoms are usually displaced away from their equilibrium positions by a small distance along the Cartesian directions. The IFCs were then determined by either finite difference scheme on the atomic forces [45] or fitting the atomic displacement-force relations, Eq. (2) [44]. Since most atoms in the supercell are staying in their equilibrium positions, the IFCs calculated from such small-displacement approaches correspond to 0 K. Here, we extract the 0 K IFCs using the fitting method.

As the temperatures increases, the amplitudes of atom displacements become larger so that the anharmonic terms in Eq. (1) cannot be treated as perturbations to the harmonic Hamiltonian. In order to apply the phonon scattering theories to determine the thermal conductivity, one possible approach is to use the harmonic Hamiltonian,

$$\widehat{H} = T + \widehat{V} = T + \widehat{E}_0 + \frac{1}{2!}\sum_{ij}\sum_{\alpha\beta} \hat{\phi}_{ij}^{\alpha\beta} u_i^\alpha u_j^\beta, \quad (3)$$

where $\hat{\phi}$ is the effective second-order harmonic force constants and $\widehat{E}_0$ is the effective equilibrium potential energy, to approximate the Hamiltonian, Eq. (1). Such an effective phonon theory has been successfully applied to one-dimensional atomic chain with high anharmoncity, where Li *et al.* [46] derived the effective phonon dispersion and group velocities, and extracted the phonon mean free paths based on these effective phonon properties. Hellaman *et al.* [34] showed that the effective phonon concept could help to obtain the phonon dispersion of high-temperature phases. The residue, $H - \widehat{H}$, are compensated by the effective anharmonic terms through

$$H - \widehat{H} = V - \widehat{V}$$
$$= E_0 - \widehat{E}_0 + \frac{1}{3!}\sum_{ijk}\sum_{\alpha\beta\gamma} \hat{\psi}_{ijk}^{\alpha\beta\gamma} u_i^\alpha u_j^\beta u_k^\gamma + \frac{1}{4!}\sum_{ijkl}\sum_{\alpha\beta\gamma\theta} \hat{\chi}_{ijkl}^{\alpha\beta\gamma\theta} u_i^\alpha u_j^\beta u_k^\gamma u_l^\theta + \cdots, \quad (4)$$

where $\hat{\psi}$ and $\hat{\chi}$ are the effective anharmonic IFCs. Since the temperature determines the range of atomic displacements, the effective harmonic Hamiltonian is temperature dependent. Correspondingly, the effective harmonic and anharmonic force constants should be temperature dependent as well.

To obtain the temperature dependent (effective) IFCs, we employ a stochastic sampling technique [36] to generate 50 different configurations corresponding to certain temperature *T*. In each configuration, 1800 carbon atoms, or equivalently 30 × 30 primitive unit cells of graphene, are included. Following Shulumba *et al.* [36], in classical system the displacements of atoms in a cell of $N_a$ atoms at $T$ should follow the distribution

$$u_i^\alpha = \sum_{s=1}^{3N_a} \varepsilon_{i\alpha}^s \langle A_{i\alpha}^s \rangle \sqrt{-2\ln\xi_1} \sin 2\pi\xi_2, \tag{5}$$

where $\varepsilon_{i\alpha}^s$ is a component that corresponds to the $\alpha$ direction of atom $i$ of the $s$-th eigenvector in the system, $\xi_1$ and $\xi_2$ are uniform random variables between (0, 1), which are used to generate normally distributed numbers. $\langle A_{i\alpha}^s \rangle$ is the thermal amplitude of normal mode from Boltzmann statistics, written as $\langle A_{i\alpha}^s \rangle = \sqrt{k_b T/m_i}/\omega_s$, where $m_i$ is the mass of atom $i$, $\omega_s$ is the frequency of mode $s$, and $k_B$ is the Boltzmann constant. With the randomly generated atomic displacements, the atomic forces can be computed based on the interatomic potential, after which the second-order effective harmonic force constants are extracted first by fitting the obtained displacement-force data through

$$F_i^\alpha = -\frac{\partial \hat{V}}{\partial u_i^\alpha} = -\sum_j \sum_\beta \hat{\phi}_{ij}^{\alpha\beta} u_j^\beta. \tag{6}$$

Then, we fit the displacement-residual force data to obtain the third-order and four-order anharmonic force constants using the relation

$$F_i^\alpha - \left(-\sum_j \sum_\beta \hat{\phi}_{ij}^{\alpha\beta} u_j^\beta\right) = -\frac{1}{2!}\sum_{jk}\sum_{\beta\gamma} \hat{\psi}_{ijk}^{\alpha\beta\gamma} u_j^\beta u_k^\gamma - \frac{1}{3!}\sum_{jkl}\sum_{\beta\gamma\theta} \hat{\chi}_{ijkl}^{\alpha\beta\gamma\theta} u_j^\beta u_k^\gamma u_l^\theta. \tag{7}$$

During the fitting processes, the translational invariances of the interatomic force constants are imposed using the singular-value decomposition technique, which is documented in Ref. [47].

It is noted that one has to know the mode frequencies and eigenvectors at temperature $T$, which are not available in advance, in order to go through the procedures mentioned above. Here, we use the harmonic force constants at 0 K as an initial guess to compute the frequencies and eigenvectors. With the displacement-force dataset generated based on the initial set of harmonic force constants, the new harmonic force constants are calculated, which are in turn to generate newer displacement-force dataset until convergence.

### B. Lattice thermal conductivity from Peierls-Boltzmann transport equation

To calculate the thermal conductivity, an isotopically pure single-layer graphene sheet is placed in the *x-y* plane. The graphene sample is sandwiched between two heat reservoirs with a distance *L* apart in the *x* direction, and the temperatures of the two reservoirs are kept at $T + \Delta T/2$ and $T - \Delta T/2$, respectively, with an average temperature of the sample $T$ and an infinitely small temperature difference $\Delta T$. The temperature gradient in the sample is $\frac{dT}{dx} = \Delta T/L$. Due to the driving of the temperature gradient, the phonon populations do not follow the equilibrium function (Bose-Einstein distribution for phonons in realistic materials and classical distribution for phonons in classical MD simulations), but are perturbed. We can write the non-equilibrium phonon populations as $n_\lambda = n_\lambda^0 + n_\lambda^0(n_\lambda^0 + 1)\frac{dT}{dx}F_\lambda^x$, where $\lambda$ ($\equiv \mathbf{q}s$) denotes the $s$-th phonon mode with a momentum $\mathbf{q}$, $n_\lambda^0$ is equilibrium phonon population, and $F_\lambda^x$ is the so-called deviation function for mode $\lambda$, which can be calculated by solving the steady-state linearized PBTE. With the solved deviation functions, the thermal conductivity along the $x$ direction is expressed by the contribution from each phonon mode

$$K^{xx} = \frac{1}{V}\sum_\lambda \hbar\omega_\lambda v_\lambda^x n_\lambda^0(n_\lambda^0 + 1)F_\lambda^x, \tag{8}$$

where $\hbar$ is the Planck's constant; $V$ represents the volume of the unit cell; $\omega_\lambda$ and $v_\lambda$ are the phonon frequency and group velocity of mode $\lambda$, respectively, which are determined by the phonon dispersion of graphene. For a defect-free graphene, the PBTE can be written as [43]

$$v_\lambda^x \frac{\partial n_\lambda^0}{\partial T}\frac{\partial T}{\partial x} = W_\lambda^{3ph} + W_\lambda^{4ph} + W_\lambda^{boundary}, \tag{9}$$

where the left-hand-side term represents the phonon diffusion driven by temperature gradient, and the right-hand side are the collision terms originated from three-phonon, four-phonon and phonon-boundary scatterings.

The phonon-phonon scattering terms on the left-hand side of Eq. (9) are expressed as [16, 29]

$$W_\lambda^{3ph} = \sum_{\lambda'}\sum_{\lambda''}\frac{\partial T}{\partial x}\left[W_{\lambda,\lambda',\lambda''}^+(F_{\lambda''}^x - F_{\lambda'}^x - F_\lambda^x) + \frac{1}{2}W_{\lambda,\lambda',\lambda''}^-(F_{\lambda''}^x + F_{\lambda'}^x - F_\lambda^x)\right], \tag{10}$$

$$W_\lambda^{4ph} = \sum_{\lambda'}\sum_{\lambda''}\sum_{\lambda'''}\frac{\partial T}{\partial x}\left[\frac{1}{2}W_{\lambda,\lambda',\lambda'',\lambda'''}^{++}(F_{\lambda'''}^x - F_{\lambda'}^x - F_{\lambda''}^x - F_\lambda^x)\right.$$

$$\left.+ \frac{1}{2}W_{\lambda,\lambda',\lambda'',\lambda'''}^{+-}(F_{\lambda''}^x + F_{\lambda'''}^x - F_{\lambda'}^x - F_\lambda^x) + \frac{1}{6}W_{\lambda,\lambda',\lambda'',\lambda'''}^{--}(F_{\lambda'}^x + F_{\lambda''}^x + F_{\lambda'''}^x - F_\lambda^x)\right] \tag{11}$$

While $W_{\lambda,\lambda',\lambda''}^\pm$ in Eq. (10) is the transition probability for the three-phonon processes $\mathbf{q}s \pm \mathbf{q}'s' \to \mathbf{q}''s''$, $W_{\lambda,\lambda',\lambda'',\lambda'''}^{\pm\pm}$ in Eq. (11) is the counterpart for the four-phonon scattering events $\mathbf{q}s \pm \mathbf{q}'s' \pm \mathbf{q}''s'' \to \mathbf{q}''s''$. Usually, the expressions of the transition probability for these phonon-phonon scatterings derived according to the Fermi golden rule are used, which are expressed as [29]

$$W^{\pm}_{\lambda,\lambda',\lambda''} = 2\pi n^0_\lambda \left(n^0_{\lambda'} + \frac{1}{2} \mp \frac{1}{2}\right)(n^0_{\lambda''} + 1)|V_3(-\lambda, \mp\lambda', \lambda'')|^2$$

$$\times \delta(\omega_\lambda \pm \omega_{\lambda'} - \omega_{\lambda''})\Delta(\mathbf{q} \pm \mathbf{q}' - \mathbf{q}'' + \mathbf{G}), \tag{12}$$

$$W^{\pm\pm}_{\lambda,\lambda',\lambda'',\lambda'''} = 2\pi n^0_\lambda \left(n^0_{\lambda'} + \frac{1}{2} \mp \frac{1}{2}\right)\left(n^0_{\lambda'} + \frac{1}{2} \mp \frac{1}{2}\right)(n^0_{\lambda'''} + 1)|V_4(-\lambda, \mp\lambda', \mp\lambda'', \lambda''')|^2$$

$$\times \delta(\omega_\lambda + \omega_{\lambda'} + \omega_{\lambda''} - \omega_{\lambda'''})\Delta(\mathbf{q} \pm \mathbf{q}' \pm \mathbf{q}'' - \mathbf{q}''' + \mathbf{G}). \tag{13}$$

$V_3$ and $V_4$ in the above equations are three-phonon and four-phonon scattering matrix elements, quantifying the strength of the scattering events. The expressions for $V_3$ and $V_4$ can be found in Refs. [29]. The momentum conservation conditions during the scattering events, i.e.,

$$\mathbf{q} \pm \mathbf{q}' - \mathbf{q}'' + \mathbf{G} = \mathbf{0}, \tag{14}$$

and

$$\mathbf{q} \pm \mathbf{q}' \pm \mathbf{q}'' - \mathbf{q}''' + \mathbf{G} = \mathbf{0}, \tag{15}$$

with a reciprocal vector $\mathbf{G}$, are ensured by the $\Delta$ function. Depending on whether $\mathbf{G}$ equals $\mathbf{0}$ or not, the scattering is a normal process or a Umklapp process. The Dirac-delta function $\delta$ in Eq. (12) and Eq. (13) means conservation of energy before and after the phonon scatterings, that is,

$$\omega_\lambda \pm \omega_{\lambda'} - \omega_{\lambda''} = 0 \tag{16}$$

and

$$\omega_\lambda \pm \omega_{\lambda'} \pm \omega_{\lambda''} - \omega_{\lambda'''} = 0, \tag{17}$$

for the three-phonon and four-phonon processes, respectively.

The phonon-boundary scattering term depends on the phonon lifetime due to phonon-boundary scattering, which is written as [48]

$$W^{boundary}_\lambda = -\frac{n_\lambda - n^0_\lambda}{\tau^{boundary}_\lambda} = -\frac{1}{a}\frac{n^0_\lambda(n^0_\lambda+1)F^x_\lambda}{L/|v^x_\lambda|}\frac{\partial T}{\partial x}. \tag{18}$$

The phonon lifetime is usually assumed to be proportional to the ratio between the sample size and the phonon group velocity. The constant number $a$ in Eq. (18) is the scaling parameter. When $L \to \infty$, $W^{boundary}_\lambda$ approaches zero and phonons in the sample only experience the phonon-phonon scatterings.

C. Numerical solution of PBTE

To calculate the thermal conductivity through Eq. (8), the first Brillouin zone of graphene is first discretized to an $N \times N$ **q**-point grid, and then the frequency and group velocity of phonon modes on the grid are calculated through the phonon dispersion relation. In order to obtain the mode-specific deviation functions, one has to solve the PBTE, a set of linear equations with respect to $\{F^x\}$. The coefficients of the linear equations are relevant to the summation of the transition probabilities of all possible phonon scattering events.

For a given mode $\lambda$ (**q**$s$) and the other two (three) phonon branches, $s'$ and $s''$ ($s'$, $s''$ and $s'''$), fixed, there is a four-dimensional (six-dimensional) space of **q**′ and **q**″ (**q**′, **q**″ and **q**‴) that can form a three-phonon (four-phonon) scattering event. Taking the advantage of the momentum conservation condition in Eqs. (14) and (15) to eliminate **q**″, the summations become over **q**′ and over **q**′ and **q**‴ for three-phonon and four-phonon scatterings, respectively.

If one treats **q**′ and **q**‴ as continuous variables, the summation over **q**′ and **q**‴ in Eqs. (10) and (11) is converted to the integral through $\sum_{\mathbf{q}'} = \frac{1}{\Omega}\int_\Omega d\mathbf{q}'$ and $\sum_{\mathbf{q}'}\sum_{\mathbf{q}'''} = \frac{1}{\Omega^2}\int_\Omega d\mathbf{q}' \int_\Omega d\mathbf{q}'''$, where $\Omega$ denotes the first Brillouin zone. Using the **q**-point grid as the quadrature grid for **q**′ and **q**‴, the above multiple integral could be determined straightforwardly through numerical integration if the integrands are continuous functions. However, additional attention should be paid to perform the above integrations when the Dirac delta function appears in the integrands. For the integration over **q**′ and **q**‴, one can always perform the numerical integration over **q**′ first by fixing **q**‴. Without loss of generality, the integrals over **q**′ in Eqs. (9) and (10) have the form

$$I(W) = \int_\Omega \phi(\mathbf{q}')\delta(\omega_{\mathbf{q}'s'} - W)d\mathbf{q}' \qquad (19)$$

where $\phi(\mathbf{q}')$ is a function with respect to **q**′.

We employ the so-called tetrahedron method to calculate the integration with the Dirac delta function [49]. In the this method, the first Brillouin zone is decomposed into a few non-overlapped tetrahedrons or triangles with the same volume or area, depending on whether the crystal is three-dimensional or two-dimensional, and both $\omega$ and $\phi$ are assumed to be linear with respect to the wavevector **q**′. When dealing with two-dimensional graphene, the integral over **q**′, Eq. (19), could be turned to a one-dimensional line integral,

$$I(W) = \int_{\omega_{\mathbf{q}'s'}=W} \phi(\mathbf{q}')/|\nabla\omega_{\mathbf{q}'s'}|dl \qquad (20)$$

As a result, the integral over this triangle is non-zero if the line of **q**′ that satisfies the energy conservation condition intersects with a specific triangle. The contribution of the line integral from each triangle can be evaluated by the values of $\omega$ and $\phi$ at the three corners of the triangle through interpolation, and the integral is written as

$$I(W) = \sum_n \sum_{m=1}^{3} r_m^n(W) \phi_n(\mathbf{q}'_m) \quad (21)$$

where $n$ and $m$ denote the $n$-th triangle and $m$-th corner point in a triangle, $a$ is the area of each triangle, $\phi_n(\mathbf{q}'_m)$ means the function value of $\phi$ at the $m$-th corner point of the $n$-th triangle. $r_m^n(W)$ is a weight factor and is related to $W$ and the corner frequency of the $s'$-th band, $\omega_{\mathbf{q}'_m s'}^n$. The expressions for $r_m^n(W)$ are provided in Appendix A.

For a crystal with finite size, the phonon frequencies are also discrete values and summations in Eqs. (10) and (11) are over a discrete set of wave vectors. For such a discrete system, it has been suggested to broaden the Dirac delta function by approximating it with a Lorentzian function [41, 50]

$$\delta(\omega) = \frac{1}{\pi} \frac{\epsilon}{\omega^2 + \epsilon^2}, \quad (22)$$

where $\epsilon$ is a small number. Turney $et\ al.$ [41] proposed to let

$$\epsilon = 2(\Gamma_{\mathbf{q}s} + \Gamma_{\mathbf{q}'s'} + \Gamma_{\mathbf{q}''s''}), \quad (23)$$

where $2\Gamma$ is phonon linewidth ($\Gamma$ is also interpreted as phonon scattering rate and $1/\Gamma$ phonon lifetime), for all three-phonon processes. Similarly, we express $\epsilon$ as

$$\epsilon = 2(\Gamma_{\mathbf{q}s} + \Gamma_{\mathbf{q}'s'} + \Gamma_{\mathbf{q}''s''} + \Gamma_{\mathbf{q}'''s'''}), \quad (24)$$

for all four-phonon processes.

Although the strict justification for the substitution of the phonon linewidth dependent Lorentzian function for the Dirac delta is not mentioned in the references [41], such a treatment is not uncommon when studying impact ionization with high scattering rates [51]. The Dirac delta function is the outcome of the Fermi golden rule, which is expected to be valid for low collision rates. When the scattering is strong, the state energies are always changing due to the self-energy shift during the process [52]. Thus, phonons whose energies do not obey Eqs. (16,17) still have the chance to make transitions occur. In Ref. [52], the expression for the scattering rates that the broadening effects are included are derived, where the so-called joint spectral density function replaces the Dirac delta. The joint spectral density function has the form [51],

$$A(E, E') = \frac{\hbar}{\pi} \frac{\Gamma(E) + \Gamma(E')}{[E + \Delta(E) - E' - \Delta(E')]^2 + [\Gamma(E) + \Gamma(E')]^2}, \quad (25)$$

where $E$ and $E'$ are the initial and final states; $\Delta$ means the energy shift. In the limit of small $\Gamma$, the golden rule is recovered.

Once the coefficients in Eqs. (10) and (11) are numerically calculated, we use an iterative method, the biconjugate gradient stabilized method [53], to self-consistently solve the set of linear equations with respect to $\{F^x\}$. Typically, ten to twenty iterations are sufficient to obtain the converged results. Using the solved $\{F^x\}$, as well as the phonon frequencies and group velocities, at the points in the **q**-grid, the thermal conductivity of graphene is obtained through Eq. (8).

D. Homogeneous nonequilibrium molecular dynamics method

Although the focus of this work is the PBTE method, we will closely compare the results from it against those from classical MD simulations. Among the various MD based methods, the homogeneous nonequilibrium MD (HNEMD) method [54, 55] is the most efficient one in the diffusive transport regime. This method was first proposed in terms of two-body potentials [54] and was recently generalized to general many-body potentials [55]. It is physically equivalent to the Green-Kubo method, but is more efficient due to the fact that the thermal conductivity is calculated directly from the heat current, instead of the heat current autocorrelation function. In the HNEMD method, one adds a driving force on top of the interatomic force for each particle and controls the system temperature during the time integration. The thermal conductivity is then simply calculated from the nonequilibrium ensemble average of the heat current. For details on this method as applied to systems described by many-body empirical potentials, see Ref. [55]. The HNEMD calculations were performed using the graphics processing units molecular dynamics (GPUMD) package [56]. The same Tersoff potential [31] as used in the PBTE calculations was adopted here. We used a time step of 1 fs in all the calculations.

**III. Results and discussion**

A. Temperature dependent interatomic force constants

Figure 1 shows the phonon dispersion of graphene along the high-symmetry lines, Γ-M-K-Γ, calculated using the harmonic force constants at 0 K and 300 K. Overall, the two sets of force constants lead to similar phonon dispersion curves, but two noticeable differences could be identified from Fig. 1(a). First, the finite temperature leads to a downshift of the phonon spectrum. For example, the obtained maximum frequency at the Γ point is reduced from 1690 cm$^{-1}$ to 1685 cm$^{-1}$ if one uses the IFCs at 300 K to replace those at 0 K. Second, as shown in Fig. 1(b), the flexural out-of-plane acoustic (ZA) phonon branch follows $\omega_{ZA} \propto q^2$ near the center of the Brillouin zone (the Γ point) if using the force constants at 0 K, but the ZA branch computed by the force constants at 300 K becomes linearized. This observation is consistent with a theoretical analysis [57], which predicts the dispersion should follow $\omega_{ZA}(q) = \alpha(T,q)q^2$ with $\alpha(T,q) = \alpha_0[1 + q_c^2/q^2]^{1/4}$ and a temperature dependent cut-off wave vector $q_c$.

The phonon shifts are originated from two sources. One is due to the thermal expansion, which is usually modeled through the quasiharmonic approximation. The thermal expansion coefficient of graphene is known to be a negative value at room temperature [58, 59], and the lattice constant of graphene that is free of stress at 300 K in MD simulation is 0.4‰ smaller than that corresponding to the minimum energy at 0 K. The deformation of the crystal could either downshift or upshift the phonon frequency depending on the mode-specific Gruneisen parameters, which are defined as $\gamma_\lambda = -(\Delta\omega_\lambda/\Delta V)/(\omega_\lambda/V)$, with the crystal volume $V$. For graphene, the Gruneisen parameters for out-of-plane acoustic modes are negative but positive for in-plane modes [60]. Thus, the span of the whole phonon spectrum is slightly increased, but the ZA modes are softened, leading to imaginary frequencies around the Γ point when the temperature is 0 K (See Fig. 1(b)). The other source for phonon shifts is the explicit anharmonicity [61]. At high temperatures, the atoms move far away from their equilibrium positions and the anharmonicity tends to make the effective potential energy surface under the harmonic approximation different from the potential energy surface near the equilibrium positions of atoms, so that the phonon dispersion becomes temperature dependent. Such a temperature effect helps to linearize the long-wavelength ZA phonons. Mariani and von Oppen [57] applied renormalization group analysis to derive the effective Hamiltonian of graphene under high temperatures and showed that the linearized ZA dispersion is related to the anharmonic coupling of the flexural modes with the in-plane modes. Since the linearizing of the ZA phonons due to the explicit anharmonicity overcomes the phonon softening due to compressive strain, the frequency of ZA modes are positive and the phonon dispersion is linear near the Γ point.

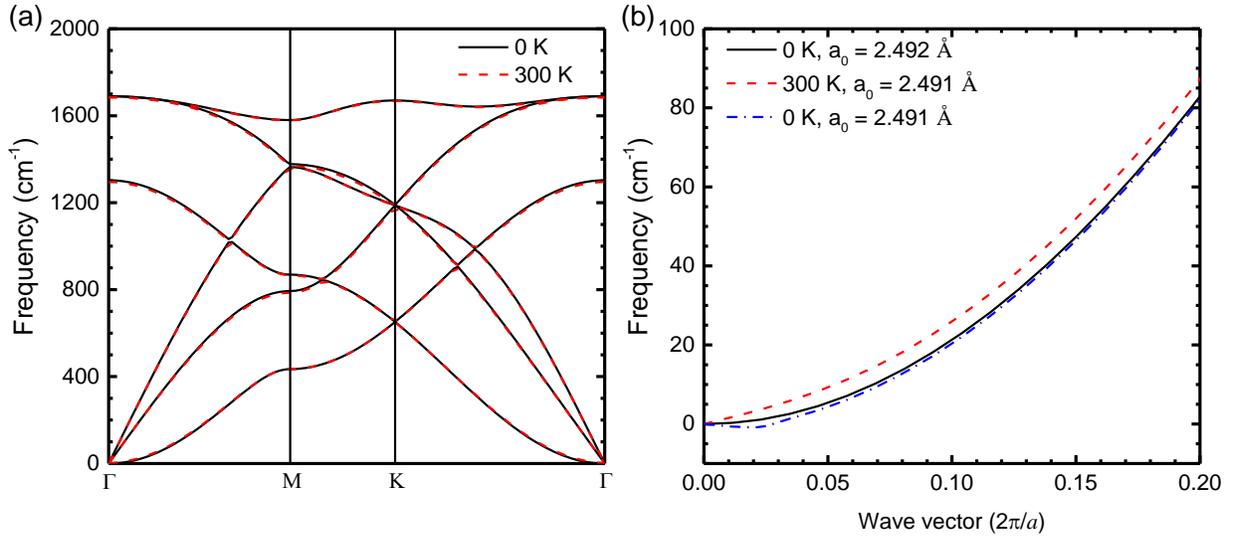

Figure 1. (a) Phonon dispersion of graphene along the high-symmetry lines, Γ-M-K-Γ. (b) Phonon dispersion curve of graphene for the ZA modes around the Γ point in the direction from Γ to M.

In addition to the second-order harmonic force constants, third-order and fourth-order anharmonic force constants are also affected by the temperature. Tables S1 and S2 of Supplemental Material list the values of the anharmonic IFCs estimated at 0 K and 300 K. Since the magnitude of the anharmonic IFCs decays rapidly with the interatomic distance, we only list $\psi_{112}^{\alpha\beta\gamma}$ and $\chi_{1112}^{\alpha\beta\gamma\theta}$, where 1 and 2 refer to two neighboring carbon atoms, as examples. The change of anharmonic IFCs due to temperature could be understood by the fact that at high temperatures the anharmonic terms in the Hamiltonian go into the harmonic terms and the effective anharmonic terms are altered. From Table S1, the temperature dependence of the third-order anharmonic force constants is found to be weak, as the change of the values of force constants is less than 10% when the temperature is elevated from 0 K to 300 K. In contrast, the fourth-order force constants, which are listed in Table S2, are heavily affected by the temperature, as seen from the two sets of data corresponding to 0 K and 300 K.

B. Thermal conductivity of graphene

Since in MD simulations all phonon modes are fully excited and the phonon population does not follow the Bose-Einstein distribution ($n_\lambda^0 = 1/\left[\exp\left(\frac{\hbar\omega_\lambda}{k_B T}\right) - 1\right]$) but the classical one ($n_\lambda^{0,C} = k_B T/\hbar\omega_\lambda$), the thermal conductivity prediction from MD simulations should be deviated from the conventional PBTE results where the quantum effects are correctly captured. Considering that the Debye temperature of graphene (~ 2000 K) is much higher than room temperature, the difference of the thermal conductivity calculated based on different statistics might be large. To obtain the classical thermal conductivity of graphene through the PBTE calculations, one might substitute the classical distribution for the quantum one directly in the PBTE. However, such a substitution leads to the breakdown of the equilibrium phonon population balance before and after the scattering. For example, the three-phonon annihilation scattering process, $\lambda + \lambda' \to \lambda''$, requires $n_\lambda n_{\lambda'}(n_{\lambda''} + 1) = (n_\lambda + 1)(n_{\lambda'} + 1)n_{\lambda''}$, which does not hold when using the classical distribution. Instead of using the simple substitution, we artificially decease the value of the Planck's constant to approach the classical limit [62], which not only makes the Bose-Einstein statistics approach to the classical one but also keeps the population balance of each phonon scattering event hold. In our calculations, the Planck's constant used is 1/100 of the original value, and we find that the thermal conductivity value of graphene at room temperature converges, with less than 1% difference if the Planck's constant is further reduced. For simplicity, we use classical and quantum thermal conductivity to refer to the thermal conductivity calculated based on the classical and Bose-Einstein distributions.

Figure 2 shows the thermal conductivity of graphene at 300 K calculated through the PBTE as a function of the number of **q**-points in the first Brillouin zone, along with the thermal conductivity prediction from equilibrium molecular dynamics (EMD) simulations with different number of unit cells [9, 10]. We additionally performed HNEMD simulations to compute the thermal conductivity of graphene, which is also plotted in Fig. 2. More detailed HNEMD results can be found in Appendix B. If the number of q-points in the PBTE calculation is similar to that of the unit cells in MD simulation, the calculated thermal conductivity values from these two methods are expected to be similar, since the same resolution of the phonon wave vectors is considered.

As most previous PBTE studies did, we first calculate the thermal conductivity using the IFCs extracted at 0 K and consider the three-phonon processes as the only phonon-phonon scattering mechanism. The analytical tetrahedron method is employed to impose the energy conservation before and after the scattering. The computed thermal conductivity is around 3100 W/mK and found to be weakly dependent on the number of q-points. Our results based on three-phonon scatterings are close to the classical thermal conductivity of graphene described by the optimized Tersoff potential from a similar three-phonon PBTE study by Singh and Fisher [19], as well as the EMD predictions from previous works [9, 10] and the HNEMD results. To understand the roles of four-phonon scatterings in graphene, we include four-phonon scattering terms in the PBTE. The obtained classical thermal conductivity is around 1200 W/mK, much smaller compared with the calculations without four-phonon scatterings. The results suggest that four-phonon scatterings should be crucial in graphene even at 300 K, but the disagreement with the EMD and HNEMD results is quite considerable. We should note that up to now the IFCs at 0 K are used in the calculations, without taking into account the effects of temperature on the IFCs and the corresponding phonon properties.

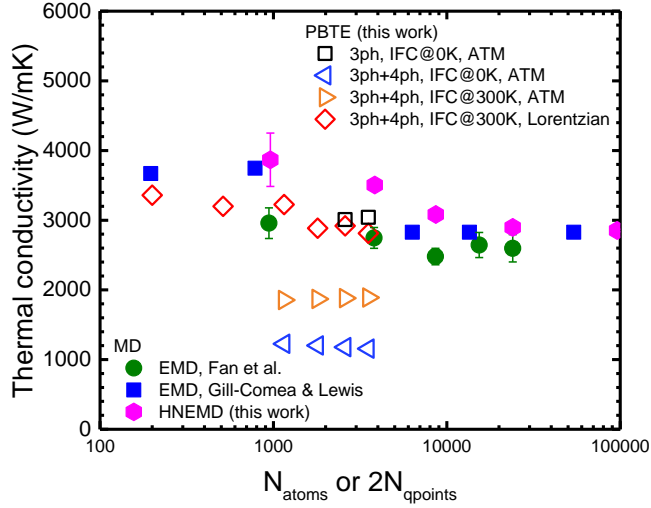

Figure 2. The computed classical thermal conductivity of graphene as a function of the number of atoms or **q**-points. "3ph" means that only three-phonon scatterings are considered in the PBTE calculations, while "3ph+4ph" indicates both three-phonon and four-phonon scatterings are included. The EMD data from Fan et al. [9], Gill-Comea and Lewis [10] and additional HNEMD results are also shown for comparison.

To incorporate the temperature effects on the IFCs, we compute the phonon dispersion and phonon scattering rates using the force constants extracted at 300 K. The scattering rate of a specific mode is related to the transition probability through

$$\Gamma_\lambda^{3\text{ph}} = \sum_{\lambda'} \sum_{\lambda''} \left( W^+_{\lambda,\lambda',\lambda''} + \frac{1}{2} W^-_{\lambda,\lambda',\lambda''} \right) / n^0_\lambda (n^0_\lambda + 1), \tag{26}$$

$$\Gamma_\lambda^{4ph} = \sum_{\lambda'} \sum_{\lambda''} \left(\frac{1}{2} W^{++}_{\lambda,\lambda',\lambda'',\lambda'''} + \frac{1}{2} W^{+-}_{\lambda,\lambda',\lambda'',\lambda'''} + \frac{1}{6} W^{--}_{\lambda,\lambda',\lambda'',\lambda'''}\right) / n_\lambda^0 (n_\lambda^0 + 1). \quad (27)$$

Figure 3(a) shows that the four-phonon scattering rates computed from the IFCs at 300 K are roughly lower by an order of magnitude compared to those obtained by using the IFCs at 0 K. However, as shown in Fig. 3(b), the three-phonon scattering rates are less affected, which could be understood by the fact that third-order anharmonic force constants do not change much with the increase of temperature from 0 K to 300 K, as discussed in Sec. III.A. Plugging these three-phonon and four-phonon scattering rates into the PBTE, we find that the obtained thermal conductivity becomes 50% larger than that based on 0 K interatomic force constants, as expected, but still lower than the MD results.

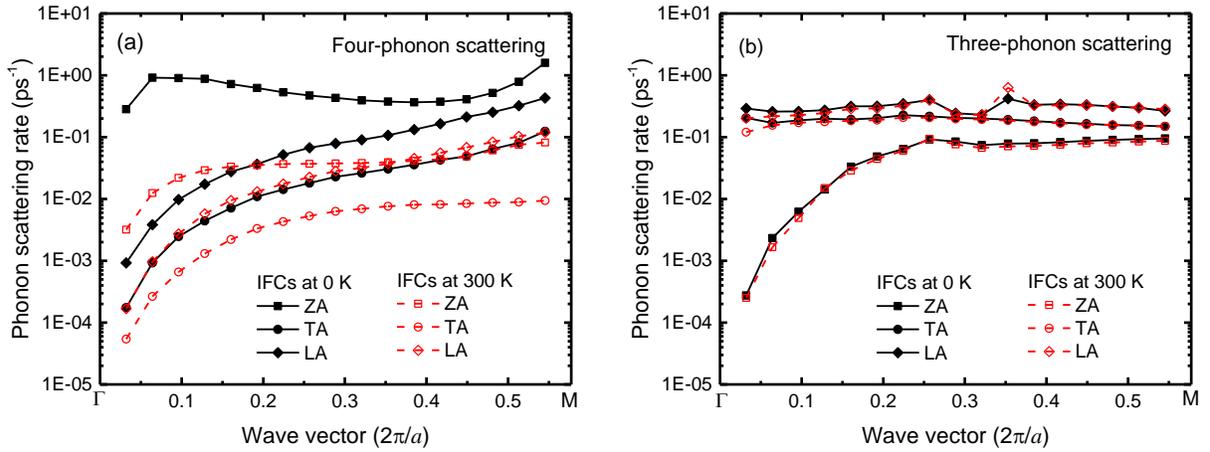

Figure 3. (a) Four-phonon and (b) three-phonon scattering rates of acoustic modes of graphene along Γ-M direction.

According to the discussion on the temperature dependent IFCs in Sec. III(A), the temperature would lead to the change of harmonic force constants, which results in phonon frequency shifts, as well as the change of anharmonic force constants, which are related to the rates of phonon scatterings as they appear in $|V_3|^2$ and $|V_4|^2$ of Eqs. (13) and (14). In order to figure out how these two different mechanisms affect phonon transport in graphene, we perform two additional PBTE calculations (the **q**-grid with $36 \times 36$ points is used) with different combinations of IFCs. The thermal conductivity is enhanced from 1180 W/mK to 1530 W/mK, when the temperature dependent harmonic force constants are employed but using the anharmonic force constants corresponding to 0 K. Using temperature dependent anharmonic and 0 K harmonic force constants leads to a thermal conductivity of 1470 W/mK. Hence, both the temperature dependence on the harmonic and anharmonic force constants are equally crucial to predict the thermal conductivity of graphene. As discussed above, three-phonon scattering rates are not significantly affected by the temperature-dependent IFCs, the change of the fourth-order IFCs should be responsible for the enhanced thermal conductivity when using the temperature dependent anharmonic IFCs. To further elucidate the roles of the temperature dependent harmonic and fourth-order IFCs on the strength of four-phonon scatterings, the four-phonon scattering rates of the ZA modes computed by using different combination of IFCs are plotted in Figure 4. When the fourth-order IFCs become temperature dependent, the percentage of phonon scattering rate downshift for different modes is roughly of the same value. Using the temperature dependent harmonic force constants also suppresses the phonon scattering rates, but the low-frequency phonons are found more likely to be affected than the high-frequency phonons. This could be attributed to the fact that the change of the anharmonic IFCs will alter the phonon scattering matrix elements, which causes the reduction of phonon scattering rates for all phonons, while the harmonic force constants are more relevant to the phonon scattering phase space, which are distinct for different modes.

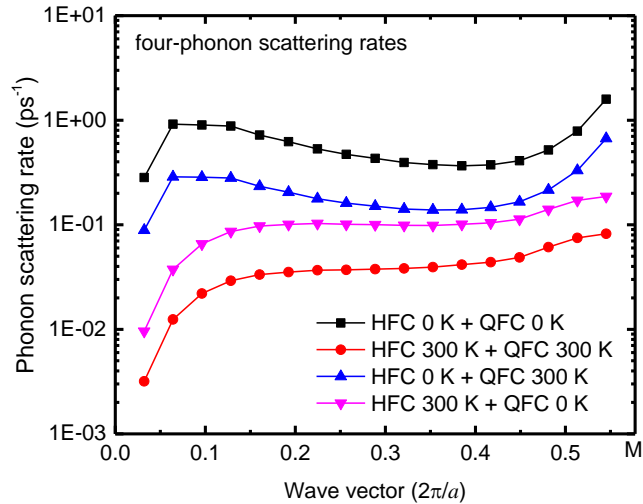

Figure 4. Phonon scattering rates of ZA modes along the Γ-M direction calculated by using different combination of the interatomic force constants.

C. Comparison with molecular dynamics simulations

To clarify the discrepancy between the results from PBTE calculations and EMD and HNEMD simulations on graphene, one has to be aware of the discrete nature of molecular dynamics. In a periodic system whose size is finite, the available wave vectors have to be $\mathbf{q} = \frac{n_1}{N_1}\mathbf{b}_1 + \frac{n_2}{N_2}\mathbf{b}_2 + \frac{n_3}{N_3}\mathbf{b}_3$ due to the Bloch's theorem, where $\mathbf{b}_i$ is the reciprocal vector, $N_i$ is the number of unit cells along the direction of the $i$-th translational vector of the crystal and $n_i$ is a non-negative integer that is smaller than $N_i$. Thus, the wave vectors and phonon frequencies are discretized in such a periodic system, and phonons have less chance to find other phonons that are of proper wave vectors and frequencies to strictly fulfill the momentum and energy conservation conditions to let the phonon scattering happen. Hence, the phonon scattering could be weaker in a finite size system compared with a larger system where more phonon modes are available. Due to the anharmonicity, phonon frequencies are broadened, which is characterized by the phonon linewidth, making the phonon scatterings happen even if the phonon frequencies do not exactly satisfy the energy conservation condition. As we will show below, the phonon broadening is crucial to correctly compute phonon scattering rates in a finite system. The justification of neglecting the phonon broadening in PBTE calculations on large crystalline materials will be discussed at the end of this section.

To take the effects of phonon broadening into account, we employ Eq. (22) to replace the Dirac delta function. Since the smearing parameter $\epsilon$ in Eq. (22) is not known before phonon linewidths are determined, we provide an initial guess for the phonon linewidths and compute them iteratively until the calculated thermal conductivity is converged within 1%. The obtained thermal conductivity data is plotted in Fig. 3 as red diamonds. When the number of q-points are 10×10, corresponding to a cell with $N_a$=200 atoms, the thermal conductivity is around 3300 W/mK. With more q-points are included in the calculations, the thermal conductivity is slightly reduced to 2800-2900 W/mK. Both thermal conductivity values and the $N_a$ dependence satisfactorily agree with the recent EMD simulations [9, 10] and our HNEMD results.

We also compute the thermal conductivity of graphene without including four-phonon scatterings and extremely high values (~30000 W/mK) are obtained with 36×36 q-points. To explore the role of three-phonon and four-phonon scatterings in a relatively small finite system, we show the phonon scattering rates (one half of the phonon linewidths) of the ZA modes in Figure

5. The extremely small three-phonon scattering rates for the low-frequency (long-wavelength) ZA modes in the finite system can be clearly identified. As is known, the main three-phonon scattering channel for the long-wavelength ZA modes is the annihilation process ZA+ZA->TA/LA [15], but typically the wavelength of the in-plane acoustic modes that participate the scattering should be longer than that of ZA modes. Since the size of the finite system is fixed, the modes whose wavelengths are larger than the dimension do not exist making the annihilation scatterings ineffective. Compared with the phonon scattering rates of a relatively small finite system, the scattering rates in a larger system, where the energy conservation condition is applied through the analytical tetrahedron method, are much larger, especially for the low-frequency modes, due to the fact that the scattering events that involve the long-wavelength in-plane mode could occur. In addition, as seen in Fig. 5, the phonon linewidths of the finite system in the three-phonon calculation is found to be smaller than those due to three-phonon scatterings in the calculations where both three-phonon and four-phonon scatterings are included, implying that four-phonon processes not only are one of the origins of phonon scatterings but also help to increase the phonon linewidth and thus to enhance three-phonon scatterings by enlarging the phonon-phonon scattering phase space. Another observation from Fig. 5 is that the four-phonon scattering rates do not change much for the finite or large graphene sheet. This is because the four-phonon scattering phase space is much larger than the three-phonon counterparts. In other words, the four-phonon scattering events that include long-wavelength modes only occupy a small portion of the entire four-phonon processes, thus the wavelength cutoff in the finite system does not alter the total four-phonon scattering rates much. Therefore, four-phonon scatterings become the main scattering channel in finite systems.

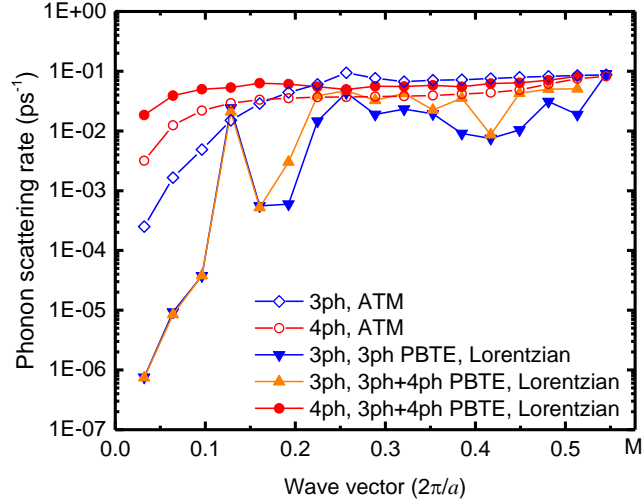

Figure 5. The calculated phonon scattering rates of the ZA modes along the Γ-M direction. "3ph" and "4ph" in the legend present the three-phonon and four-phonon scattering rates, respectively. "3ph PBTE" and "3ph+4ph PBTE" represent whether or not four-phonon scatterings are included in the PBTE calculations.

To further confirm that the PBTE calculations with phonon linewidth dependent Lorentzian function as the Dirac delta could well reproduce MD predictions, we shows the length dependent (classical) thermal conductivity of graphene at 300 K calculated through the PBTE in Fig. 6, along with the thermal conductivity prediction from non-equilibrium molecular dynamics (NEMD) simulations [6-8]. The length $L$ for both PBTE and NEMD data is the distance between the hot and cold reservoirs. Following Mingo and Broido [48], the constant, $a$, in the phonon-boundary term, Eq. (17), is chosen to be 2.0. The choice has been widely employed in PBTE calculations to model phonon-boundary scatterings [16, 63, 64], since it is expected to recover the ballistic phonon transport regime automatically in the small $L$ limit. One the one hand, plugging Eq. (17) into Eq. (8) and neglecting the phonon-phonon scattering terms in the small $L$ limit, one can easily solve the PBTE and obtain

$$n_\lambda = n_\lambda^0 + \frac{v_{qs}^x}{|v_{qs}^x|} \frac{\Delta T}{2} \frac{dn_{qs}^0}{dT}. \qquad (27)$$

One the other hand, in the ballistic limit the population function of phonons traveling to the right (left) equals the equilibrium population function of phonons emitted by the left (right) thermal reservoir. At the middle of the sample, $x = L/2$, for a right-going (left-going) mode $\lambda$, one has $n_\lambda = n_\lambda^0 + (-)\frac{dn_\lambda^0}{dT}\frac{\Delta T}{2}$, which is identical to the solution of the PBTE, Eq. (8). When the sample size is smaller than 30 nm, the classical thermal conductivity from our PBTE calculations, which is shown as the black solid line, agrees well with the NEMD data (hollow symbols), as well as the results from atomistic Green's function simulations [13], where phonon transport is assumed to be ballistic and the thermal conductance of graphene, $G = K/L$, is found to be around 10 GW/m²K when the phonon population distribution approaches classical at high temperature limit.

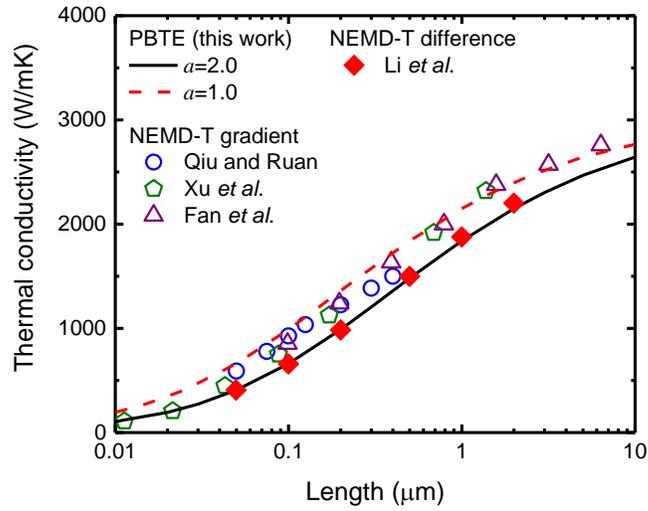

Figure 6. Classical thermal conductivity of graphene at 300 K as a function of sample size. The data from NEMD simulations in which the temperature gradient is determined by fitting the linear temperature profile in the sample are represented by open symbols: Qiu and Ruan [8], Xu *et al.* [6] and Fan *et al.* [7]. The results of NEMD simulations by Li *et al.* [65] where the temperature gradient is determined by the temperature difference of the two reservoirs are shown as red diamonds.

However, with the increase of the sample length, the PBTE results become smaller than the prediction from NEMD simulations. This observation is in accordance with the study by Tadano *et al.* [38], where they found that the PBTE calculations with $a = 2.0$ tend to severely underestimate the thermal conductivity of silicon compared with the NEMD data. They identified that $a = 1.0$ could well reproduce the size dependence of the thermal conductivity of bulk silicon, but did not further explain the success of $a = 1.0$. Inspired by the finding from Tadano *et al.* [38], we also perform calculations using $a = 1.0$ in the phonon-boundary scattering term, and the results are shown as red solid line in Fig. 6. In the sample length range from 50 nm to 10 μm, the thermal conductivity from the PBTE calculations becomes larger than the results corresponding to $a = 2.0$, and close to the prediction from the NEMD simulations.

A possible explanation for the failure of $a = 2.0$ for large $L$ to reproduce the NEMD data is as follows. In these NEMD simulations, the thermal conductivity is computed through $K = -J/(\partial T/\partial x)$, where $J$ is the heat flux and $\partial T/\partial x$ is the temperature gradient determined by fitting the linear temperature profile in the middle region of the sample, so that the effects of phonon-boundary scatterings on phonon transport between the two reservoirs are not correctly considered. To recognize this issue, one might think about a purely harmonic crystalline material sandwiched between two reservoirs. There is no temperature gradient within the sample but the thermal resistance occurs at the edges between the sample and the reservoirs. Using the linear temperature region to calculate the temperature gradient, the thermal conductivity becomes infinitely large. However, since phonons experience ballistic transport, the thermal conductivity is not infinite but a sample-size dependent value. Therefore, the reported thermal conductivity from most NEMD studies should be overestimated and one has to use a weaker phonon-boundary scattering ($a = 1.0$) in PBTE calculations to match the NEMD results. Very recently, some researchers suggested to use $\frac{\partial T}{\partial x} = \Delta T/L$ to determine the temperature gradient in NEMD calculations [65], which is expected to reproduce the ballistic thermal conductivity in weak anharmonic materials. In Fig. 6, We plot the thermal conductivity data from their NEMD simulations as red diamonds. This treatment makes the length dependent thermal conductivity from NEMD simulations excellently agree with our PBTE calculations.

Despite the good agreement between PBTE calculations and MD simulations, one may ask whether the thermal conductivity as well as phonon scattering rates computed using the two different methods of treating the energy conservation are converged when the number of q-points is sufficiently large. To explain this, let us consider a one-dimensional system and assume that the q-points grid is so dense that the phonon dispersion can be treated as quasi-continuous. The integral involving the $\delta$ function, Eq. (18), can be written as

$$I(W) = \int_\Omega \phi(\mathrm{q}') \frac{\eta}{\left(\omega_{\mathrm{q}'s'}-W\right)^2 + \eta^2} d\mathrm{q}', \tag{28}$$

$$I(W) = \int_\Omega \phi(\mathrm{q}') \frac{\epsilon}{\left(\omega_{\mathrm{q}'s'} - W\right)^2 + \epsilon^2} d\mathrm{q}', \tag{29}$$

where $\eta$ is an infinitely small number and $\epsilon$ a finite number that is determined through Eqs. (23) and (24), for the Dirac delta and Lorentzian functions, respectively. For a region near $\mathrm{q}'_0$ where $\omega_{\mathrm{q}'_0 s'} = W$, its contribution to the integral is $\phi(\mathrm{q}'_0)/|\partial \omega_{\mathrm{q}'_0 s'}/\partial \mathrm{q}'|$ for the Dirac delta. Meanwhile, for the Lorentzian function, one could find a neighboring region near $\mathrm{q}'_0$ out of which the contribution to the integral is negligible. The size of the region positively depends on $\epsilon$, or correspondingly the phonon linewidths of the material. If $\epsilon$ is small, the variation of $\phi(\mathrm{q}')$ in the small region would be negligible as well, resulting in the same results as in Eq. (28). In contrast, if the phonon linewidths are large, which occur at high temperatures and highly anharmonic crystals, the two approaches dealing with the $\delta$ function should give different results. Since the phonon linewidths of graphene is short due to the weak anharmonicity in graphene, one should expect that converged thermal conductivity is achieved when using dense q-grid.

### D. Temperature dependent thermal conductivity

Figure 7 shows the temperature dependent thermal conductivity of graphene. With the inclusion of four-phonon scatterings and the phonon linewidth dependent Lorentzian function as the Dirac delta function, the calculated classical thermal conductivity of graphene using temperature dependent IFCs again agrees decently with our HNEMD simulations.

We also compute both classical and quantum thermal conductivity using the tetrahedron method, in which the wave vectors of graphene are assumed continuous. When both three-phonon and four-phonon scatterings are considered, the difference between classical and quantum thermal conductivity is found to be less than 10% even at 300 K. Although 300 K is still much lower than the Debye temperature, the small difference is expected due to a cancellation between two competing factors: 1) the mode heat capacity of high-frequency phonons is overestimated in the classical system; 2) the phonons become more likely to be scattered since more high-frequency phonons are available to take part in the phonon-phonon scatterings in the classical system. In the whole temperature range we explored, the classical thermal conductivity is slightly lower than the quantum thermal conductivity, suggesting that compared with the Bose-Einstein statistics the reduction of the thermal conductivity due to stronger phonon scatterings in the classical system overweighs the enhancement due to the change of heat capacity.

To quantify the importance of four-phonon scatterings, we examine the ratio $(K^{3\text{ph}} - K^{4\text{ph}})/K^{4\text{ph}}$. At 300 K, neglecting four-phonon scatterings would lead to a 90% overestimation of the quantum thermal conductivity. With the elevation of the temperature to 900 K, the ratio increases to 115%, confirming that four-phonon processes are more significant at higher temperatures [27, 29]. The crucial role of four-phonon scatterings at high temperatures is due to that the four-phonon scattering rates roughly follow the scaling relation $T^2$ while the three-phonon counterparts exhibit a $T$ dependence.

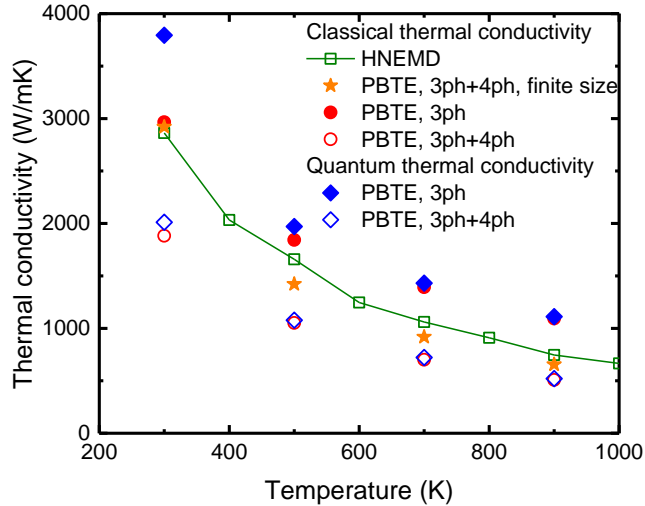

Figure 7. Thermal conductivity of graphene as a function of temperature.

### IV. Summary and conclusions

In summary, we study the phonon scattering mechanisms and thermal conductivity in single-layer graphene, whose interatomic interaction is described by the optimized Tersoff potential, using the PBTE formalism. Two different methods are applied to determine the IFCs of graphene. The IFCs corresponding to 0 K are extracted by the widely used small-displacement method, and temperature dependent IFCs are obtained through fitting the displacement-force data from MD simulations. The 0 K IFCs are found to substantially underestimate the thermal conductivity of graphene by predicting much stronger four-phonon scatterings compared with the calculations based on temperature dependent IFCs. Furthermore, we show that in order to reproduce the thermal conductivity of graphene from MD simulations the discrete nature of phonon modes in the system and phonon broadening due to anharmonicity has to be considered when determining the phonon scattering rates. By using the phonon linewidth dependent Lorentzian function to approximate the Dirac delta function, which represents energy conservation condition before and after phonon scatterings, satisfactory agreement between MD simulations and PBTE calculations is achieved. The PBTE formalism employed here could also be applied to explore phonon transport in other two-dimensional and high-temperature functional materials.

**Acknowledgements**

X. G. acknowledges the support from the National Science Foundations of China (NSFC) (No. 51706134) and the Shanghai Pujiang Program (No. 17PJ1404500). Z. F. acknowledges the support from the Academy of Finland Centre of Excellence program QTF (Project 312298) and the computational resources provided by Aalto Science-IT project and Finland's IT Center for Science (CSC). H. B. acknowledges the support from the National Science Foundations of China (NSFC) (No. 51676121). The calculations were performed at the High Performance Computing Center of Shanghai Jiao Tong University.

**Appendix A. Weight coefficients of tetrahedron methods for two-dimensional crystal**

Suppose $\omega_{\mathbf{q}_1's'}^n < \omega_{\mathbf{q}_2's'}^n < \omega_{\mathbf{q}_3's'}^n$. If $\omega_{\mathbf{q}_1's'}^n < W < \omega_{\mathbf{q}_2's'}^n$, the expressions for $r_m^n$ in Eq. (21) are given by

$$r_1^n = \frac{(\omega_{\mathbf{q}_2's'}^n - W)(-\omega_{\mathbf{q}_1's'}^n + W)}{2\left(\omega_{\mathbf{q}_2's'}^n - \omega_{\mathbf{q}_1's'}^n\right)^2 \left(\omega_{\mathbf{q}_3's'}^n - \omega_{\mathbf{q}_1's'}^n\right)} + \frac{(\omega_{\mathbf{q}_3's'}^n - W)(-\omega_{\mathbf{q}_1's'}^n + W)}{2\left(\omega_{\mathbf{q}_2's'}^n - \omega_{\mathbf{q}_1's'}^n\right)\left(\omega_{\mathbf{q}_3's'}^n - \omega_{\mathbf{q}_1's'}^n\right)^2}, \tag{30}$$

$$r_2^n = \frac{\left(-\omega_{\mathbf{q}_1's'}^n + W\right)^2}{2\left(\omega_{\mathbf{q}_2's'}^n - \omega_{\mathbf{q}_1's'}^n\right)^2 \left(\omega_{\mathbf{q}_3's'}^n - \omega_{\mathbf{q}_1's'}^n\right)}, \tag{31}$$

$$r_3^n = \frac{\left(-\omega_{\mathbf{q}_1's'}^n + W\right)^2}{2\left(\omega_{\mathbf{q}_2's'}^n - \omega_{\mathbf{q}_1's'}^n\right)\left(\omega_{\mathbf{q}_3's'}^n - \omega_{\mathbf{q}_1's'}^n\right)^2}. \tag{32}$$

When $\omega_{\mathbf{q}_2's'}^n < W < \omega_{\mathbf{q}_3's'}^n$, the expressions become

$$r_1^n = \frac{\left(\omega_{\mathbf{q}_3's'}^n - W\right)^2}{2\left(\omega_{\mathbf{q}_3's'}^n - \omega_{\mathbf{q}_1's'}^n\right)^2 \left(\omega_{\mathbf{q}_3's'}^n - \omega_{\mathbf{q}_2's'}^n\right)}, \tag{33}$$

$$r_2^n = \frac{\left(\omega_{\mathbf{q}_3's'}^n - W\right)^2}{2\left(\omega_{\mathbf{q}_3's'}^n - \omega_{\mathbf{q}_1's'}^n\right)\left(\omega_{\mathbf{q}_3's'}^n - \omega_{\mathbf{q}_2's'}^n\right)^2}, \tag{34}$$

$$r_3^n = \frac{(-\omega_{\mathbf{q}_1's'}^n + W)(\omega_{\mathbf{q}_3's'}^n - W)}{2\left(\omega_{\mathbf{q}_3's'}^n - \omega_{\mathbf{q}_1's'}^n\right)^2\left(\omega_{\mathbf{q}_3's'}^n - \omega_{\mathbf{q}_2's'}^n\right)} + \frac{(\omega_{\mathbf{q}_3's'}^n - W)(-\omega_{\mathbf{q}_2's'}^n + W)}{2\left(\omega_{\mathbf{q}_3's'}^n - \omega_{\mathbf{q}_1's'}^n\right)\left(\omega_{\mathbf{q}_3's'}^n - \omega_{\mathbf{q}_2's'}^n\right)^2}. \tag{35}$$

**Appendix B. Detailed thermal conductivity results from HNEMD calculations**

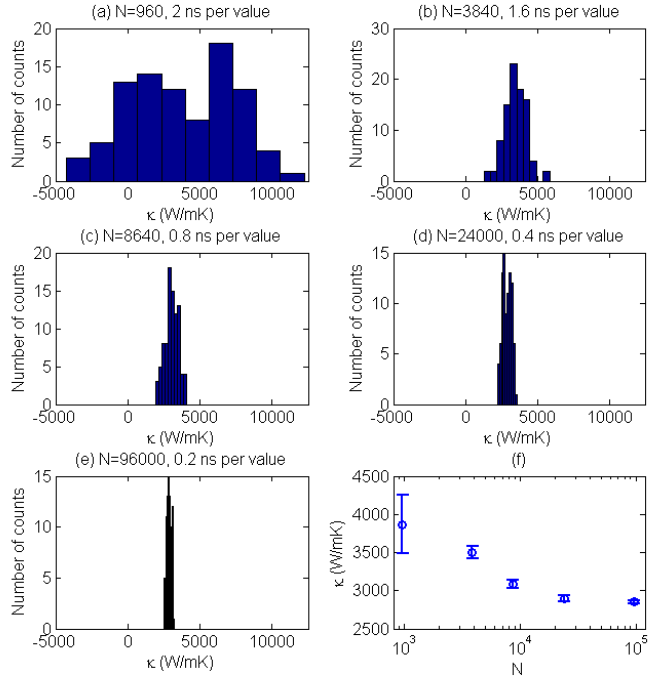

Fig 8. (a)-(e) Distribution (number of counts) of the thermal conductivity values from the HNEMD method. From (a) to (e), the numbers of atoms in the simulation cell are respectively 960, 3840, 8640, 24000, and 96000, the numbers of independent simulations (each with a production time of 18 ns in steady state) are 9, 8, 4, 2, and 1, and each thermal conductivity value is calculated from a time interval of 1.8, 1.6, 0.8, 0.4, and 0.2 ns. Therefore, there are 90 thermal conductivity values for each simulation cell size. (f) The averaged thermal conductivity, with the error bar calculated as the standard error of the distribution, as a function of the number of atoms in the simulation cell. One can see that in the HNEMD method, a larger system requires a shorter simulation time to achieve a given statistical accuracy. This is another advantage of the HNEMD method over the Green-Kubo method.

**References**

[1] A.A. Balandin, S. Ghosh, W. Bao, I. Calizo, D. Teweldebrhan, F. Miao, C.N. Lau, Superior thermal conductivity of single-layer graphene, Nano letters 8(3) (2008) 902-907.

[2] W. Cai, A.L. Moore, Y. Zhu, X. Li, S. Chen, L. Shi, R.S. Ruoff, Thermal transport in suspended and supported monolayer graphene grown by chemical vapor deposition, Nano letters 10(5) (2010) 1645-1651.

[3] A.A. Balandin, Thermal properties of graphene and nanostructured carbon materials, Nature materials 10(8) (2011) 569.

[4] E. Pop, V. Varshney, A.K. Roy, Thermal properties of graphene: Fundamentals and applications, MRS bulletin 37(12) (2012) 1273-1281.

[5] H. Bao, J. Chen, X. Gu, B.-Y. Cao, A Review of Simulation Methods in Micro/Nanoscale Heat Conduction, ES Energy & Environment 1 (2018) 16-55.

[6] X. Xu, L.F. Pereira, Y. Wang, J. Wu, K. Zhang, X. Zhao, S. Bae, C.T. Bui, R. Xie, J.T. Thong, Length-dependent thermal conductivity in suspended single-layer graphene, Nature communications 5 (2014) 3689.

[7] Z. Fan, L.F.C. Pereira, P. Hirvonen, M.M. Ervasti, K.R. Elder, D. Donadio, T. Ala-Nissila, A. Harju, Thermal conductivity decomposition in two-dimensional materials: Application to graphene, Physical Review B 95(14) (2017) 144309.

[8] B. Qiu, Y. Wang, Q. Zhao, X. Ruan, The effects of diameter and chirality on the thermal transport in free-standing and supported carbon-nanotubes, Applied Physics Letters 100(23) (2012) 233105.

[9] Z. Fan, L.F.C. Pereira, H.-Q. Wang, J.-C. Zheng, D. Donadio, A. Harju, Force and heat current formulas for many-body potentials in molecular dynamics simulations with applications to thermal conductivity calculations, Physical Review B 92(9) (2015) 094301.

[10] M. Gill-Comeau, L.J. Lewis, Heat conductivity in graphene and related materials: A time-domain modal analysis, Physical Review B 92(19) (2015) 195404.

[11] M. Park, S.-C. Lee, Y.-S. Kim, Length-dependent lattice thermal conductivity of graphene and its macroscopic limit, Journal of Applied Physics 114(5) (2013) 053506.

[12] Y. Lu, J. Guo, Thermal transport in grain boundary of graphene by non-equilibrium Green's function approach, Applied Physics Letters 101(4) (2012) 043112.

[13] A.Y. Serov, Z.-Y. Ong, E. Pop, Effect of grain boundaries on thermal transport in graphene, Applied Physics Letters 102(3) (2013) 033104.

[14] G. Fugallo, A. Cepellotti, L. Paulatto, M. Lazzeri, N. Marzari, F. Mauri, Thermal conductivity of graphene and graphite: collective excitations and mean free paths, Nano letters 14(11) (2014) 6109-6114.

[15] N. Bonini, J. Garg, N. Marzari, Acoustic phonon lifetimes and thermal transport in free-standing and strained graphene, Nano letters 12(6) (2012) 2673-2678.

[16] X. Gu, R. Yang, First-principles prediction of phononic thermal conductivity of silicene: A comparison with graphene, Journal of Applied Physics 117(2) (2015) 025102.

[17] L. Lindsay, D. Broido, N. Mingo, Flexural phonons and thermal transport in graphene, Physical Review B 82(11) (2010) 115427.

[18] L. Lindsay, W. Li, J. Carrete, N. Mingo, D. Broido, T. Reinecke, Phonon thermal transport in strained and unstrained graphene from first principles, Physical Review B 89(15) (2014) 155426.

[19] D. Singh, J.Y. Murthy, T.S. Fisher, Mechanism of thermal conductivity reduction in few-layer graphene, Journal of Applied Physics 110(4) (2011) 044317.

[20] Y. Kuang, L. Lindsay, S. Shi, X. Wang, B. Huang, Thermal conductivity of graphene mediated by strain and size, International Journal of Heat and Mass Transfer 101 (2016) 772-778.

[21] D. Broido, M. Malorny, G. Birner, N. Mingo, D. Stewart, Intrinsic lattice thermal conductivity of semiconductors from first principles, Applied Physics Letters 91(23) (2007) 231922.

[22] A. Ward, D. Broido, D.A. Stewart, G. Deinzer, Ab initio theory of the lattice thermal conductivity in diamond, Physical Review B 80(12) (2009) 125203.

[23] W. Li, J. Carrete, N.A. Katcho, N. Mingo, ShengBTE: A solver of the Boltzmann transport equation for phonons, Computer Physics Communications 185(6) (2014) 1747-1758.

[24] K. Esfarjani, G. Chen, H.T. Stokes, Heat transport in silicon from first-principles calculations, Physical Review B 84(8) (2011) 085204.

[25] X. Gu, R. Yang, Phonon transport and thermal conductivity in two-dimensional materials, Annual Review of Heat Transfer 19 (2016) 1-65.

[26] X. Gu, Y. Wei, X. Yin, B. Li, R. Yang, Colloquium: Phononic thermal properties of two-dimensional materials, Reviews of Modern Physics 90 (2018).

[27] T. Feng, X. Ruan, Quantum mechanical prediction of four-phonon scattering rates and reduced thermal conductivity of solids, Physical Review B 93(4) (2016) 045202.

[28] X. Gu, C. Zhao, Thermal conductivity of hexagonal Si, Ge, and Si1-xGex alloys from first-principles, Journal of Applied Physics 123(18) (2018) 185104.

[29] T. Feng, L. Lindsay, X. Ruan, Four-phonon scattering significantly reduces intrinsic thermal conductivity of solids, Physical Review B 96(16) (2017) 161201.

[30] T. Feng, X. Ruan, Four-phonon scattering reduces intrinsic thermal conductivity of graphene and the contributions from flexural phonons, Physical Review B 97(4) (2018) 045202.

[31] L. Lindsay, D. Broido, Optimized Tersoff and Brenner empirical potential parameters for lattice dynamics and phonon thermal transport in carbon nanotubes and graphene, Physical Review B 81(20) (2010) 205441.

[32] J.M. Skelton, L.A. Burton, S.C. Parker, A. Walsh, C.-E. Kim, A. Soon, J. Buckeridge, A.A. Sokol, C.R.A. Catlow, A. Togo, Anharmonicity in the High-Temperature C m c m Phase of SnSe: Soft Modes and Three-Phonon Interactions, Physical review letters 117(7) (2016) 075502.

[33] K. Parlinski, Z. Li, Y. Kawazoe, First-principles determination of the soft mode in cubic ZrO 2, Physical Review Letters 78(21) (1997) 4063.

[34] O. Hellman, P. Steneteg, I.A. Abrikosov, S.I. Simak, Temperature dependent effective potential method for accurate free energy calculations of solids, Physical Review B 87(10) (2013) 104111.

[35] O. Hellman, D.A. Broido, Phonon thermal transport in Bi 2 Te 3 from first principles, Physical Review B 90(13) (2014) 134309.

[36] N. Shulumba, O. Hellman, A.J. Minnich, Intrinsic localized mode and low thermal conductivity of PbSe, Physical Review B 95(1) (2017) 014302.

[37] A. Maradudin, A. Fein, Scattering of neutrons by an anharmonic crystal, Physical Review 128(6) (1962) 2589.

[38] T. Tadano, Y. Gohda, S. Tsuneyuki, Anharmonic force constants extracted from first-principles molecular dynamics: applications to heat transfer simulations, Journal of Physics: Condensed Matter 26(22) (2014) 225402.

[39] D. Broido, A. Ward, N. Mingo, Lattice thermal conductivity of silicon from empirical interatomic potentials, Physical Review B 72(1) (2005) 014308.

[40] G. Fugallo, M. Lazzeri, L. Paulatto, F. Mauri, Ab initio variational approach for evaluating lattice thermal conductivity, Physical Review B 88(4) (2013) 045430.

[41] J. Turney, E. Landry, A. McGaughey, C. Amon, Predicting phonon properties and thermal conductivity from anharmonic lattice dynamics calculations and molecular dynamics simulations, Physical Review B 79(6) (2009) 064301.

[42] A.K. Vallabhaneni, D. Singh, H. Bao, J. Murthy, X. Ruan, Reliability of Raman measurements of thermal conductivity of single-layer graphene due to selective electron-phonon coupling: A first-principles study, Physical Review B 93(12) (2016) 125432.

[43] G.P. Srivastava, The physics of phonons, CRC press1990.

[44] K. Esfarjani, H.T. Stokes, Method to extract anharmonic force constants from first principles calculations, Physical Review B 77(14) (2008) 144112.

[45] X. Tang, B. Fultz, First-principles study of phonon linewidths in noble metals, Physical Review B 84(5) (2011) 054303.

[46] N. Li, P. Tong, B. Li, Effective phonons in anharmonic lattices: Anomalous vs. normal heat conduction, EPL (Europhysics Letters) 75(1) (2006) 49.

[47] X. Gu, X. Li, R. Yang, Phonon transmission across Mg 2 Si/Mg 2 Si 1− x Sn x interfaces: A first-principles-based atomistic Green's function study, Physical Review B 91(20) (2015) 205313.

[48] N. Mingo, D. Broido, Length dependence of carbon nanotube thermal conductivity and the "problem of long waves", Nano letters 5(7) (2005) 1221-1225.

[49] P.E. Blöchl, O. Jepsen, O.K. Andersen, Improved tetrahedron method for Brillouin-zone integrations, Physical Review B 49(23) (1994) 16223.

[50] A.J. Ladd, B. Moran, W.G. Hoover, Lattice thermal conductivity: A comparison of molecular dynamics and anharmonic lattice dynamics, Physical Review B 34(8) (1986) 5058.

[51] J. Bude, K. Hess, G. Iafrate, Impact ionization in semiconductors: Effects of high electric fields and high scattering rates, Physical Review B 45(19) (1992) 10958.

[52] D.K. Ferry, Quantum transport in semiconductors, Semiconductors, Macmillan, New York, 1991.

[53] H.A. Van der Vorst, Bi-CGSTAB: A fast and smoothly converging variant of Bi-CG for the solution of nonsymmetric linear systems, SIAM Journal on scientific and Statistical Computing 13(2) (1992) 631-644.

[54] D.J. Evans, Homogeneous NEMD algorithm for thermal conductivity—application of non-canonical linear response theory, Physics Letters A 91(9) (1982) 457-460.

[55] Z. Fan, H. Dong, A. Harju, T. Ala-Nissila, Homogeneous nonequilibrium molecular dynamics method for heat transport and spectral decomposition with many-body potentials, Physical Review B 99(6) (2019) 064308.

[56] Z. Fan, W. Chen, V. Vierimaa, A. Harju, Efficient molecular dynamics simulations with many-body potentials on graphics processing units, Computer Physics Communications 218 (2017) 10-16.

[57] E. Mariani, F. von Oppen, Flexural phonons in free-standing graphene, Physical review letters 100(7) (2008) 076801.

[58] D. Yoon, Y.-W. Son, H. Cheong, Negative thermal expansion coefficient of graphene measured by Raman spectroscopy, Nano letters 11(8) (2011) 3227-3231.

[59] N. Bonini, M. Lazzeri, N. Marzari, F. Mauri, Phonon anharmonicities in graphite and graphene, Physical review letters 99(17) (2007) 176802.

[60] N. Mounet, N. Marzari, First-principles determination of the structural, vibrational and thermodynamic properties of diamond, graphite, and derivatives, Physical Review B 71(20) (2005) 205214.

[61] G. Grimvall, Thermophysical properties of materials, Elsevier1999.

[62] J. Li, Modeling microstructural effects of deformation resistance and thermal conductivity, Massachusetts Institute of Technology, 2000.

[63] H. Xie, T. Ouyang, É. Germaneau, G. Qin, M. Hu, H. Bao, Large tunability of lattice thermal conductivity of monolayer silicene via mechanical strain, Physical Review B 93(7) (2016) 075404.

[64] L. Lindsay, D. Broido, Enhanced thermal conductivity and isotope effect in single-layer hexagonal boron nitride, Physical Review B 84(15) (2011) 155421.

[65] Z. Li, S. Xiong, C. Sievers, Y. Hu, Z. Fan, N. Wei, H. Bao, S. Chen, D. Donadio, T. Ala-Nissila, Influence of Boundaries and Thermostatting on Nonequilibrium Molecular Dynamics Simulations of Heat Conduction in Solids, unpublished.